\documentclass[12pt]{article}

\usepackage{amsmath}
\usepackage{amssymb}
\usepackage{graphicx}

\begin{document}

\title{Dynamic critical phenomena from\\ spectral functions on the lattice}

\author{
J{\"u}rgen Berges$^1$, S{\"o}ren Schlichting$^1$, and D{\'e}nes Sexty$^{1,2}$
\\[0.5cm]
$^1$Institute for Nuclear Physics\\
Darmstadt University of Technology\\
Schlossgartenstr. 9, 64289 Darmstadt, Germany \\
\\
$^2$Institute for Theoretical Physics \\
University of Heidelberg \\
Philosophenweg 16, 69120 Heidelberg, Germany}
\date{}
\begin{titlepage}
\maketitle
\def\thepage{}          % No page number on title page

\begin{abstract}
We investigate spectral functions in the vicinity of the critical
temperature of a second-order phase transition. Since critical
phenomena in quantum field theories are governed by classical
dynamics, universal properties can be computed using real-time lattice
simulations.  For the example of a relativistic single-component
scalar field theory in $2+1$ dimensions, we compute the spectral
function described by universal scaling functions and extract the
dynamic critical exponent $z$. Together with exactly known static
properties of this theory, we obtain a verification from first
principles that the relativistic theory is well described by the
dynamic universality class of relaxational models with conserved
density (Model C).
\end{abstract}
 
\end{titlepage}

\renewcommand{\thepage}{\arabic{page}}

\section{Introduction}

Spectral functions play an important role in our understanding of
dynamic properties of matter in extreme conditions, ranging from
heavy-ion collisions and physics of the early universe to laboratory
experiments of ultracold quantum gases. In the context of collision
experiments of heavy nuclei, crucial questions about which are the
relevant dynamical degrees of freedom near and above the QCD
transition are encoded in spectral functions. Likewise, transport
properties such as described by shear and bulk viscosities can be
obtained from the slope of appropriate spectral functions at small
energies. Though these transport coefficients are computed in thermal
equilibrium, they contain important information about time-dependent
properties and provide essential ingredients for hydrodynamic
descriptions.

Approximate resummation techniques have been successfully implemented
to study even far-from-equilibrium behavior and thermalization of
spectral functions in scalar~\cite{Aarts:2001qa} and fermionic quantum
field theories~\cite{Berges:2002wr} so far. Simulations from first
principles on a real-time lattice might be based on stochastic
quantization techniques~\cite{Berges:2005yt}, but these are still in
its infancies. Lattice simulations are typically performed in
Euclidean space-time, and the reconstruction of real-time quantities
based on analytic continuation using bayesian methods such as
MEM~\cite{Asakawa:2000tr} is often difficult.

A particular challenge concerns dynamic critical phenomena near
continuous phase transitions, such as the QCD critical point in the
phase diagram of temperature versus baryon number density. At the
transition point anomalously large fluctuations signal scale invariant
physics, where also a hydrodynamic description based on a separation
of scales becomes inapplicable. A standard procedure for the
description of dynamic critical phenomena is to identify the relevant
hydrodynamic degrees of freedom away from the critical temperature with
the associated dynamic universality
class~\cite{Hohenberg:1977ym}. This approach successfully explains a
large body of experiments in condensed matter physics, and the same
may be expected for applications to relativistic heavy-ion
experiments~\cite{Son:2004iv} though a rigorous justification remains
an interesting problem. 

In this work we discuss the use of classical-statistical simulation
techniques on a real-time lattice to compute the spectral function in
the vicinity of a second-order phase transition. The universal
properties of quantum field theories can be rigorously simulated from
a classical counterpart in the same universality class. This is a
well-established fact for static properties and we explain why it also
holds for dynamic properties. The method extends an earlier suggestion
to employ the fluctuation-dissipation relation or Kubo-Martin-Schwinger 
periodicity condition~\cite{KMS} to
compute spectral functions from averages of products of classical
fields~\cite{Aarts:2001yx}. Non-equilibrium dynamics of systems near 
second-order phase transitions using classical fields were also investigated 
in~\cite{Borsanyi:2001rc,Rajantie:1999mp}. As an example, we compute the spectral
function of a relativistic single-component scalar field theory in two
spatial dimensions. We extract universal scaling functions and the
dynamic scaling exponent $z$. Together with exactly known static
properties of this theory, we obtain a verification from first
principles that the relativistic theory is well described by the
dynamic universality class of relaxational models with conserved 
density (Model C) in the classification scheme of Halperin and
Hohenberg~\cite{Hohenberg:1977ym}.

Though only universal properties will exactly agree for the quantum
and the classical theory, we also investigate the non-universal
behavior in a range of temperatures around the critical temperature
$T_c$ for which classical-statistical fluctuations are expected to
dominate over quantum corrections. We observe that for temperatures of
$T \gtrsim 1.5\, T_c$ the spectral function is approximately described by a
broad Breit-Wigner form, where the non-zero "quasi-particle" mass and
width are of the same order of magnitude. At around ten to twenty
percent deviation from $T_c$ a "twin-peak" structure appears due to
additional strong contributions from the zero-frequency domain of
critical phenomena. At $T_c$ we find that the quasi-particle peak is
suppressed and the spectral function obeys a power-law behavior at
small frequencies and momenta.

\section{Spectral function}

For an arbitrary bosonic Heisenberg operator $\hat{O}(t,{\mathbf x})$
the spectral function can be defined from the expectation value of the
commutator as
\begin{equation}
\rho(t-t', {\mathbf x}-{\mathbf x}', T) \, = \, i \langle [ \hat{O}(t,{\mathbf x}),\hat{O}^\dagger(t',{\mathbf x}') ] \rangle \, .
\label{eq:rhodef}
\end{equation}
This expectation value in thermal equilibrium at a temperature $T$ is obtained from
\begin{equation}
\langle \hat{O}(t,{\mathbf x}) \rangle \, = \, 
\frac{1}{Z}\, \mathrm{Tr}\left( e^{-\hat{H}/T} \hat{O}(t,{\mathbf x}) \right) \quad , \quad Z \, = \, \mathrm{Tr}\, e^{-\hat{H}/T} \, ,
\label{eq:expectation}
\end{equation}
where the trace is taken over the Hilbert space for given Hamilton operator $\hat{H}$. We also define the so-called statistical two-point function from the anti-commutator: 
\begin{equation}
F(t-t', {\mathbf x}-{\mathbf x}', T) \, = \, \frac{1}{2}\, \langle \{ \hat{O}(t,{\mathbf x}),\hat{O}^\dagger(t',{\mathbf x}') \} \rangle \, 
- \langle \hat{O}(t, {\mathbf x}) \rangle \langle \hat{O}^\dagger(t', {\mathbf x'}) \rangle \, .
\label{eq:Fdef}
\end{equation}
In thermal equilibrium the spectral and statistical two-point functions are connected by the fluctuation-dissipation relation or the so-called "KMS" condition in Fourier space~\cite{Berges:2004yj}:
\begin{equation}
F(\omega,p,T) \, = \, -i \left( n_T(\omega) + \frac{1}{2} \right) \rho(\omega,p,T) \, .
\label{eq:flucdiss}
\end{equation}
The Bose-Einstein distribution $n_T(\omega) = 1/(\exp(\omega/T) - 1)$ depends only on the frequency $\omega$ for given temperature, and we denote the spatial momentum as $p \equiv |{\mathbf p}|$. In $d$ spatial dimensions the Fourier transform reads 
\begin{equation}
F(\omega, p, T) \, =\, \int {\mathrm d}t\, {\mathrm d}^d x \, e^{i(\omega t - {\mathbf p}{\mathbf x})} F(t,\mathbf{x},T) \, .
\label{eq:FFourier}
\end{equation}

As an example, we will consider a relativistic real single-component scalar field theory. For the one-particle spectral and statistical functions $\hat{O}(t,\mathbf{x}) = \hat{O}^\dagger(t,\mathbf{x}) = \hat{\varphi}(t,\mathbf{x})$ with Heisenberg field operator $\hat{\varphi}(t,\mathbf{x})$. In mean field type approximations the spectral function (\ref{eq:rhodef}) would then be
\begin{equation}
\rho_0(\omega, p, T) \, =\, 2\pi i\, \mathrm{sgn}(\omega)\, \delta\!\left( 
\omega^2 - p^2 - M^2(T) \right)\, ,
\label{eq:rhofree}
\end{equation}
where $M(T)$ denotes an effective, temperature dependent mass. In the vicinity of the critical temperature $T_c$ of a second-order phase transition the mean field behavior for the mass is given by $M^2(T) \sim |T-T_c|$. In this case one observes that (\ref{eq:rhofree}) obeys the scaling behavior
\begin{equation}
\rho_0(s \omega, s p, s^2 |T-T_c|) \, =\, s^{-2} \, \rho_0(\omega, p, |T-T_c|) \, ,
\label{eq:rhofreecrit}
\end{equation}
where $s$ denotes a real, positive scale parameter. The spectral function (\ref{eq:rhofree}) 
describes propagating modes with linear dispersion relation. We will study in detail below the emergence 
of slow, diffusive modes near the phase transition, which affect the scaling behavior
significantly.

\section{Critical dynamics}

For a discussion of the physics near the critical temperature $T_c$ of a second-order phase transition it is convenient to introduce the reduced temperature
\begin{equation}
T_r \, \equiv \, \frac{T-T_c}{T_c} \, ,
\label{eq:reducedT}
\end{equation}
where we will consider $T_r \geq 0$ if not stated otherwise.
In the limit $T \to T_c$, for $\omega \to 0$ and $p \to 0$ the spectral function can exhibit scaling behavior:\footnote{Here we consider $\omega \geq 0$, $t \geq 0$ and, due to spatial isotropy, $p \equiv |\mathbf{p}|$.} 
\begin{equation}
\rho(s^z \omega, sp, s^\frac{1}{\nu}\, T_r) \, = \, s^{-(2-\eta)}\, \rho(\omega,p,T_r) \, .
\label{eq:scalingrho}
\end{equation}
The critical exponent $\eta$ is the anomalous dimension, $\nu$ is the correlation-length exponent and $z$ denotes the dynamic scaling exponent. Comparison with (\ref{eq:rhofreecrit}) shows that in mean field approximations these would be $\eta = 0$, $\nu = 1/2$, and $z = 1$ from this quantity for the relativistic field theory. Dynamic universality classes can be identified, using that at sufficiently large distances and time scales the dynamics is described by the slowest varying degrees of freedom. In hydrodynamic approaches the low-energy effective theory includes, apart from the order parameter, the densities of conserved charges~\cite{Hohenberg:1977ym}. If couplings to the latter are properly taken into account this leads to a mean field value $z=2$ as will be discussed in the conclusions.

From (\ref{eq:scalingrho}) we recover the well-known limiting cases. For instance, because the physics is scale-invariant, we can choose the scale parameter such that $s^z \omega = 1$. In this case one finds from (\ref{eq:scalingrho}) for $p=0$ at $T=T_c$:
\begin{equation}
\rho(\omega,0,0) \, \sim \, \omega^{-\frac{2-\eta}{z}} \, .
\label{eq:omegacrit}
\end{equation}
We will frequently consider (\ref{eq:scalingrho}) Fourier transformed with respect to time:
\begin{equation}
\rho(s^{-z} t, sp, s^\frac{1}{\nu}\, T_r) \, = \, s^{-(2-z-\eta)}\, \rho(t,p,T_r) \, .
\label{eq:tTscalingrho}
\end{equation}
For instance, for $p=0$ this leads with $s^{-z} t = 1$ to
\begin{equation}
\rho(t,0,T_r) \, = \, t^{\frac{2-\eta}{z}-1}\, g\left( \frac{t}{\xi_t} \right)\, ,
\label{eq:gtTscalingrho}
\end{equation}
with the scaling function $g(t/\xi_t) \equiv \rho(1,0,(t/\xi_t)^{1/(\nu z)})$. Here we have used the temporal correlation length $\xi_t$, which scales with the reduced temperature as 
\begin{equation}
\xi_t \, \sim \, T_r^{-\nu z}  \, .
\label{eq:tcorrelationlength}
\end{equation} 
The scaling function is expected to behave as
\begin{equation}
g\left( \frac{t}{\xi_t} \right) \, \sim \, e^{-\frac{t}{\xi_t}}\, .
\label{eq:scalingfunction}
\end{equation}
Since the temporal correlation length (\ref{eq:tcorrelationlength}) diverges at $T=T_c$ showing critical slowing down, the scaling function (\ref{eq:scalingfunction}) approaches a constant and according to (\ref{eq:gtTscalingrho}) the spectral function obeys a power-law behavior in time. 

For a non-zero temperature phase transition, where $T_c > 0$, the limit $\omega \to 0$ implies $\omega \ll T_c$. As a consequence, in this case the fluctuation dissipation relation (\ref{eq:flucdiss}) can be replaced by
\begin{eqnarray}
F(\omega, p, T_r) \, \stackrel{\omega \ll T}{\to} \, - i \, \frac{T}{\omega}\, \rho(\omega, p, T_r) \, .
\label{eq:clflucdiss}
\end{eqnarray}
Here the Bose-Einstein distribution leads to the classical distribution $n_T(\omega) \sim T/\omega$ for $\omega \ll T$. Equation (\ref{eq:clflucdiss}) becomes exact in the limit $\omega \to 0$ relevant for critical behavior, which leads with (\ref{eq:scalingrho}) to the corresponding scaling behavior for the statistical two-point function: 
\begin{equation}
F(s^z \omega, sp, s^\frac{1}{\nu}\, T_r) \, = \, s^{-(2-\eta+z)}\, F(\omega,p,T_r) \, .
\label{eq:scalingF}
\end{equation}
Equation (\ref{eq:clflucdiss}) can be written after a temporal Fourier transform
\begin{equation}
\rho(t,p, T_r) \, \to \, - \frac{1}{T} \, \partial_t F(t,p, T_r) \, .
\label{eq:KMStime}
\end{equation}
Here the low-frequency limit for the validity of (\ref{eq:clflucdiss}) in the quantum theory translates into the long-time behavior of the correlators in the vicinity of $T_c$.

\section{Real-time lattice simulations}

The fluctuation-dissipation relation, given by (\ref{eq:flucdiss}) for a bosonic quantum theory, relates the spectral and statistical functions in thermal equilibrium. It is an identity, which can be used to determine whether a theory is governed by quantum or classical statistics in terms of the occupation number distribution $n_T(\omega)$. While for a bosonic quantum theory $n_T(\omega)$ corresponds to the Bose-Einstein distribution, a classical-statistical bosonic field theory is characterized by (\ref{eq:clflucdiss}) with $n_T(\omega) = T / \omega$ for all frequencies. According to (\ref{eq:clflucdiss}) this classical condition is fulfilled in the quantum theory only for small frequencies, $\omega \ll T$. Here it is important to note that only momenta in the limit $\omega \to 0$, $p \to 0$ contribute to universal quantities. As a consequence, critical phenomena at a non-zero temperature phase transition are rigorously characterized by classical dynamics.

We employ this in the following to obtain the dynamic scaling exponent $z$ as well as universal scaling functions for the spectral function from classical-statistical lattice simulations. In a classical theory there are no commutation relations and the statistical two-point function corresponding to (\ref{eq:Fdef}) becomes~\cite{Berges:2004yj} 
\begin{equation}
F_\mathrm{cl}(t-t', {\mathbf x}-{\mathbf x}', T) \, = \, \langle O(t,{\mathbf x}) O^\dagger(t',{\mathbf x}') \rangle_\mathrm{cl} \, 
- \left\langle O(t, {\mathbf x}) \right\rangle_\mathrm{cl} \left\langle O(t', \mathbf{x}') \right\rangle_\mathrm{cl} \, .
\label{eq:Fcldef}
\end{equation}
We consider a classical Hamiltonian $H(\varphi,\pi)$ of the scalar field $\varphi(t,\mathbf{x})$ and its conjugate momentum field $\pi(t,\mathbf{x})$. The statistical average in (\ref{eq:Fcldef}) is then given by
\begin{equation}
\langle O(t,{\mathbf x}) O^\dagger(t',{\mathbf x}') \rangle_\mathrm{cl} \, = \, 
\frac{1}{Z_\mathrm{cl}} \int \left[\mathrm{d} \pi_0 \right] \left[ \mathrm{d} \varphi_0 \right] e^{- H/T} O(t,{\mathbf x}) O^\dagger(t',{\mathbf x}')\, ,
\label{eq:claverage}
\end{equation}
with the classical partition function $Z_\mathrm{cl} = \int \left[\mathrm{d} \pi_0 \right] \left[ \mathrm{d} \varphi_0 \right] e^{- H/T}$. The functional integral is over classical phase-space with $\int\left[\mathrm{d} \pi_0 \right] \left[ \mathrm{d} \varphi_0 \right] = \int \Pi_\mathbf{x} \mathrm{d} \pi_0(\mathbf{x}) \mathrm{d} \varphi_0(\mathbf{x})$ for given fields $\pi_0(\mathbf{x}) = \pi(t=0,\mathbf{x})$ and
$\varphi_0(\mathbf{x}) = \varphi(t=0,\mathbf{x})$ at some arbitrary initial time $t=0$. The time dependence of $O(t,{\mathbf x}) = O[\varphi(t,{\mathbf x}),\pi(t,{\mathbf x})]$ is then obtained from Hamilton's equations of motion for the fields $\varphi(t,\mathbf{x})$ and $\pi(t,\mathbf{x})$. For known $F_\mathrm{cl}(t , {\mathbf x}, T)$ the classical-statistical spectral function $\rho_\mathrm{cl}(t , {\mathbf x}, T)$ can then be efficiently obtained from~\cite{Aarts:2001yx}
\begin{equation}
\rho_\mathrm{cl}(t,p, T) \, = \, - \frac{1}{T} \, \partial_t F_\mathrm{cl}(t,p, T) \, .
\label{eq:clKMStime}
\end{equation}
Alternatively, a direct computation of the spectral function in the classical theory would involve the more complicated computation of Poisson brackets~\cite{Berges:2004yj}. 
Comparison of (\ref{eq:clKMStime}) with (\ref{eq:KMStime}) shows that near the critical temperature of a second-order phase transition the classical theory describes the long-time behavior of the corresponding quantum theory. This universal asymptotic behavior is characterized by critical exponents and scaling functions as described above.

\section{Critical dynamics in 2+1 dimensions}

As an example, we compute the one-particle spectral function of a relativistic scalar field theory with one component in two spatial dimensions. It is described by the Hamiltonian
\begin{equation}
H \, = \, \int \mathrm{d}^2 x \left( \frac{1}{2}\, \pi^2 + \frac{1}{2} (\nabla \varphi)^2 + \frac{1}{2}\, m^2 \varphi^2 + \frac{\lambda}{4 !}\, \varphi^4 \right) \, , 
\label{eq:hamiltonian}
\end{equation}
with real mass-parameter $m^2 < 0$ and coupling $\lambda > 0$ such that
$\lambda/|m| = 1$ without loss of generality for universal
properties.  This theory has served as a paradigm for the
understanding of second-order phase transitions at non-vanishing
temperature~\cite{Onsager}. Together with exactly known static
properties of this theory, we will use the example to obtain a
verification from first principles that the relativistic theory is
well described by the dynamic universality class of relaxational
models with energy conservation (Model C) in the classification scheme
of Halperin and Hohenberg~\cite{Hohenberg:1977ym}.

We extract the dynamic scaling exponent $z$ and universal scaling
functions from real-time lattice simulations. For this we define the
field theory on a spatial lattice with $N \times N$ sites with periodic
boundary conditions and lattice spacing $a$. Universal long-distance
properties do not depend on the lattice spacing, but the finite volume
has to be taken into account. We consider lattices with $N=1024$,
$256$ and $128$ to verify whether convergence of the results is
obtained for given $T_r$, $p$ and $\omega$ (or $t$).  To generate
thermal initial conditions we use a Metropolis algorithm, while the
subsequent real-time evolution of $\rho_\mathrm{cl}(t,p, T)$ is
obtained from Hamiltonian dynamics for the fields. We
typically use $30$--$160$ independently thermalized initial configurations
for each temperature. For the Hamiltonian evolution part, we average over
results with different initial but same relative time along the
trajectory using time translation invariance. This increases 
the effective number of independent configurations by a factor of 
 $\sim \Delta t /\xi_t$, where
$\Delta t$ is a sampling time of typically $1$--$3$ $10^5 a$. 

With $\pi(t, \mathbf{x}) = \dot{\varphi}(t, \mathbf{x})$ the scalar field equation of 
motion reads\footnote{Replacing in (\ref{eq:classeom}) the second-order time derivative
with a first-order term leads to a time-dependent Ginzburg-Landau equation (Model A)~\cite{Hohenberg:1977ym}.}
\begin{equation}
\ddot \varphi (t, \mathbf{x}) \, = \, \left( \bigtriangleup - m^2 \right) \varphi(t, \mathbf{x}) - \frac{\lambda}{6}\, \varphi\, ^3 (t, \mathbf{x}). 
\label{eq:classeom}
\end{equation}
For the numerical implementation of the equation the leap frog  
discretization is used on a cubic space-time lattice, with spatial and temporal lattice spacings $a$, $a_t$ to give:
\begin{eqnarray}
\tilde{\varphi}(t+a_t, \mathbf{x}) &=& 2 \tilde{\varphi} (t, \mathbf{x}) - \tilde{\varphi}(t-a_t, \mathbf{x}) - \, \tilde{m}^2 \frac{a_t^2}{a^2} \tilde{\varphi} (t, \mathbf{x}) - \frac{\tilde{\lambda}}{6} \frac{a_t^2}{a^2}  \tilde{\varphi}^3(t, \mathbf{x})
\nonumber\\ 
&& + \, \frac{a_t^2}{a^2} \sum _{i=1}^2 
\left( \tilde{\varphi} (t,\mathbf{x}+a \hat{i}) - 2 \tilde{\varphi} (t, \mathbf{x}) + \tilde{\varphi}(t, \mathbf{x} - a \hat{i}) \right)  \, , 
\label{eq:discreteclass} 
\end{eqnarray}
where $ \tilde{\varphi}(t,\mathbf{x}) = \sqrt{a} \varphi(t,\mathbf{x}) $, $\tilde{m}^2 = m^2 a^2$, and $\tilde{\lambda} = \lambda a$ is the dimensionless 
field variable, mass squared, and coupling, respectively. Here $\hat{i}$ is the unit vector in the spatial direction indicated by $i$. We solve this equation using $m^2 a^2 = -1$ and $\lambda\, a=1$ with typically $a_t/a= 0.1$.

Since the aim is to compute dynamical behavior, we use the exactly
known static exponents for the two-dimensional Ising universality
class~\cite{Onsager}. Taking the spatial correlation length exponent $\nu = 1$ and
the anomalous dimension $\eta = 1/4$, other static exponents are
related to the two independent ones in terms of scaling relations such
as $\delta = 4/\eta-1$ leading to $\delta = 15$. We will verify
further below as a check that our simulations reproduce well values
of static exponents. For this, we will also couple the order
parameter to a small external source $B$, adding a term $H_B = - \int
\mathrm{d}^2 x B \varphi$ to the Hamiltonian (\ref{eq:hamiltonian}). This is also used to numerically obtain an approximate value of the critical temperature from the vanishing of the order parameter $\phi(T,B) = \langle
\varphi(t,\mathbf{x}) \rangle_\mathrm{cl} = 0$ when $T \to T_c$ is approached from below
and $B \to 0$. From this we 
find $T_c/|m| = 4.45 \pm 0.05 $ for $N=128$. To obtain more accurate estimates, an efficient procedure is to determine the temperature at which the real-time behaviour 
of the spectral function exhibits a power-law at sufficiently late times for $B = 0$, as is explained in more detail below.
On the $N=1024$ lattice this gives $ T_c / |m| = 4.444 \pm 0.005. $
If not stated otherwise we will consider $B = 0$ and $T \ge T_c$ in the
following.

\begin{figure}[t]
\begin{center}
\includegraphics[scale=1.0]{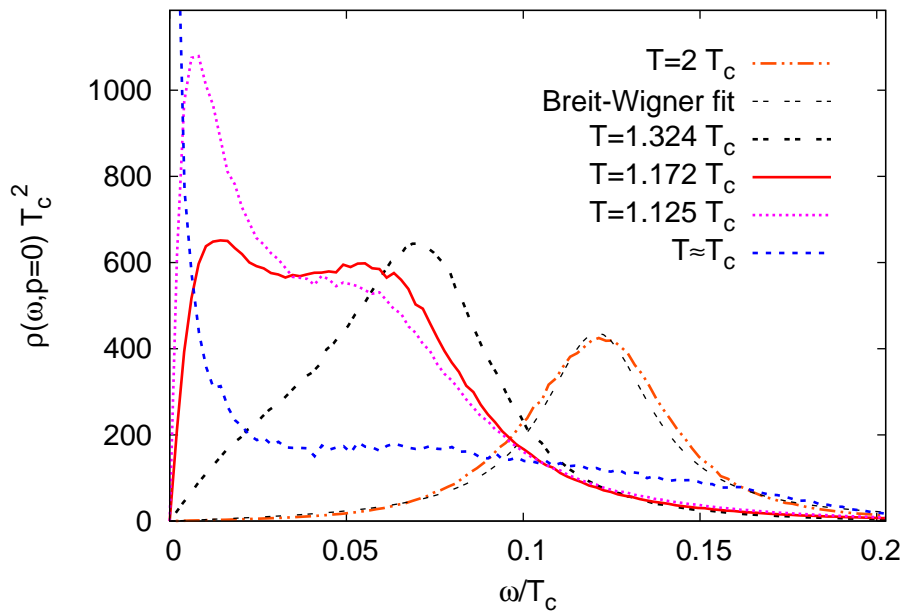}
\caption{The spectral function as a function of $\omega$ at different temperatures for $p=0$. The approximate Breit-Wigner form at $T = 2\, T_c$ goes over to a  
"twin-peak" structure while an infrared divergence builds up as $T \to T_c$.
\label{fig:twinpeak}}
\end{center}
\end{figure}
Using (\ref{eq:clKMStime}) and (\ref{eq:Fcldef}) the spectral function is given by
\begin{equation}
\rho_\mathrm{cl}(t,p, T) \, = \, - \frac{1}{T} \int {\mathrm d}^d x \, e^{-i \mathbf{p} (\mathbf{x}-\mathbf{y})} \,\langle \pi(t,\mathrm{x}) \varphi(0,\mathrm{y}) \rangle_\mathrm{cl} 
\, .
\label{eq:clrhotime}
\end{equation}    
Since thermal equilibrium is time translation invariant, a particular choice of the initial time $t=0$ in (\ref{eq:clrhotime}) is irrelevant. Fig.~\ref{fig:twinpeak} shows the spectral function (\ref{eq:clrhotime}) for different temperatures as a function of frequency $\omega$ at momentum $p = 0$. For $T = 2\, T_c$ its shape is approximately described by a broad Breit-Wigner curve whose quasi-particle peak position is at $\omega_0/T_c \simeq 0.12$ while the width is
$\gamma_0/T_c=0.036$. This means that the peak appears already for this temperature at comparably small frequencies. As a consequence, for these frequency values the Bose-Einstein distribution is well described by $n_{T}(\omega) \simeq T/\omega$ and the classical theory is expected to give an accurate description. Of course, results for the quantum theory can only be rigorously inferred in the immediate vicinity of $T_c$ in the limit of vanishing $\omega$ and $p$. 

As the temperature decreases, the peak position moves to smaller $\omega$. However, instead of a single peak continuously approaching zero near $T_c$, a second peak appears before. At temperatures of about ten to twenty percent above $T_c$ a twin-peak structure is visible from Fig.~\ref{fig:twinpeak}. Even at only about one percent deviation from the critical temperature, a remnant of the quasi-particle peak still remains visible next to the strong peak anticipating the divergence in the zero-frequency domain of critical phenomena. At $T_c$ the spectral function becomes non-analytic around $\omega = 0$: Because of the anti-symmetry property $\rho_\mathrm{cl}(- \omega,p, T) = - \rho_\mathrm{cl}(\omega,p, T)$ the spectral function vanishes at $\omega = 0$, while it diverges to $\pm \infty$ approaching this point from above and below.  

\begin{figure}[t]
\begin{center}
\includegraphics[scale=1.0]{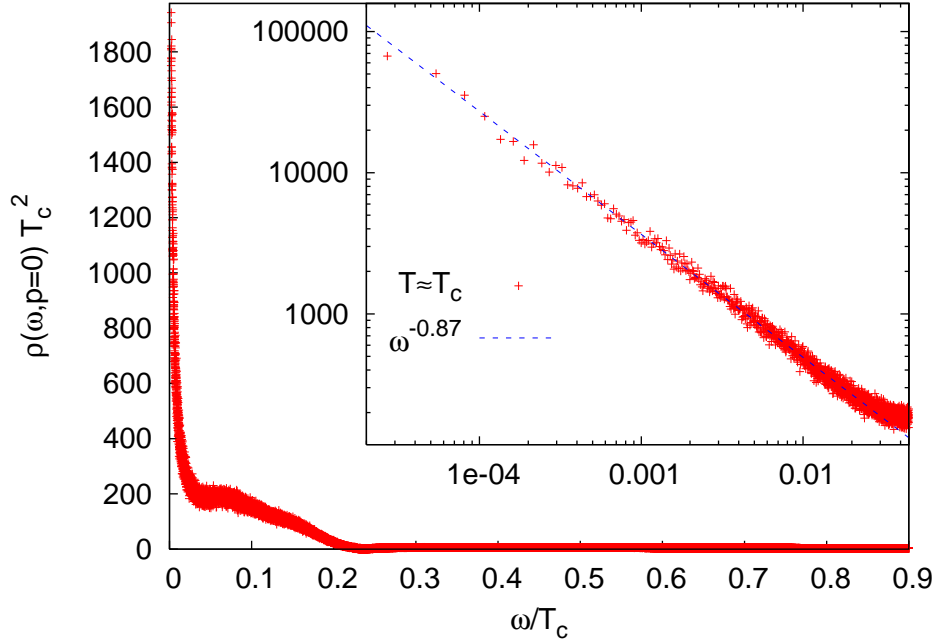}
\caption{The critical spectral function data without connecting lines as in Fig.~\ref{fig:twinpeak}. The double-logarithmic inset exhibits
a power-law in the infrared. \label{fig:inset}}
\end{center}
\end{figure}
In Fig.~\ref{fig:inset} the nature of this divergence in the vicinity of $T_c$ becomes apparent from the double-logarithmic plot in the inset. The data points exhibit a power-law behavior towards low frequencies ranging over several orders of magnitude in $\omega$. Comparison of the fit value for the exponent with (\ref{eq:omegacrit}) leads to $(2-\eta)/z \simeq 0.87$ and with $\eta = 1/4$ to the following estimate for the dynamic scaling exponent:
\begin{equation}
z \, = \, 2.0 \pm 0.1 \, .
\label{eq:z}
\end{equation}      
Besides the statistical error, we also take into account the error in 
the determination of the critical temperature. 

\begin{figure}[t]
\begin{center}
\includegraphics[scale=1.0]{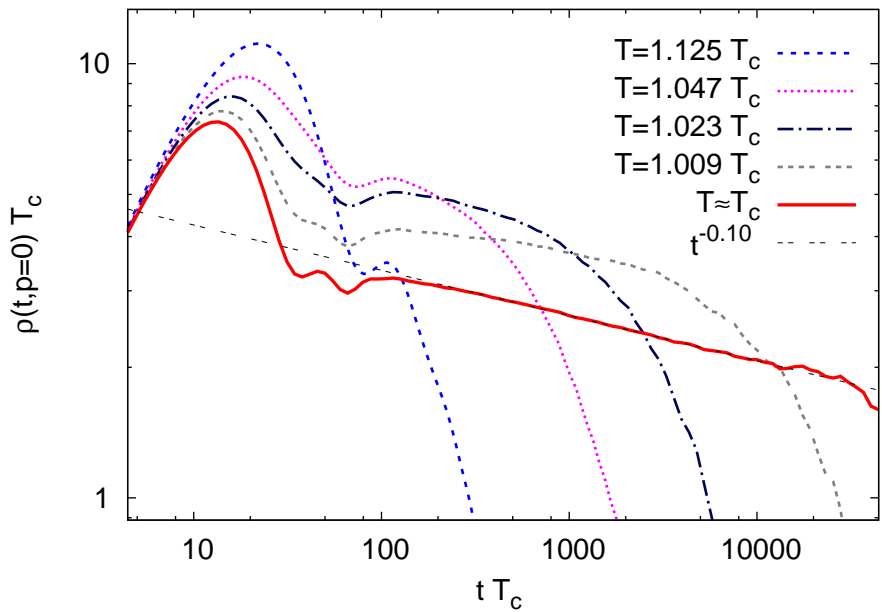}
\caption{The spectral function as a function of time $t$ at different temperatures for $p=0$. Near $T_c$ the short-time oscillatory behavior is followed by a power-law, before an exponential sets in for $t \gtrsim \xi_t$. \label{fig:plot12}}
\end{center}
\end{figure}
The corresponding temporal evolution of the spectral function is
displayed in Fig.~\ref{fig:plot12} for different temperatures and $p =
0$. According to (\ref{eq:gtTscalingrho}) and
(\ref{eq:scalingfunction}) one expects at $T_c$ a power-law behavior
in time for diverging temporal correlation length $\xi_t$. If there is
a small deviation from $T_c$ then $\xi_t$ stays large but finite and
an exponential decay will always appear for sufficiently large times
$t \gg \xi_t$. This characterization of the dynamics with respect to
power-law and/or exponential behavior neglects possible oscillatory
parts on short-time scales, which arise from structure of the spectral
function in Fourier space at larger frequencies such as remnants of
quasi-particle peaks seen in Figs.~\ref{fig:twinpeak} and
\ref{fig:inset}. Universal properties can only be extracted after the
oscillatory short-time physics is damped out. Sufficiently close to
$T_c$ the three different stages are visible in Fig.~\ref{fig:plot12}:
For $T \simeq T_c$ the damped oscillations at short times are followed
by a power-law behavior, which lasts here about two orders of
magnitude in $t T_c$, until at sufficiently late times an exponential
decay takes over. Since an infinite correlation length cannot be
achieved on a finite computer, the power-law may be extracted before
the exponential dominates. The data at $T \simeq T_c$ together with
(\ref{eq:gtTscalingrho}) yields $1-(2-\eta)/z \simeq \, 0.10$ --
$0.11$. With $\eta = 1/4$ this leads to an estimate for $z$ consistent
with (\ref{eq:z}) to two significant digits.  Going further away from
$T_c$ the temporal correlation length $\xi_t$ decreases such that the
exponential behavior of the scaling function $g(t/\xi_t)$ of
(\ref{eq:scalingfunction}) quickly dominates. Accordingly,
Fig.~\ref{fig:plot12} exhibits after the oscillatory behavior an
exponential decay without a clean power-law in between.

So far, we have extracted information about the dynamic scaling
exponent, which is obtained by putting all parameters to their
critical values ($T=T_c$, $p=0$) except one that is varied ($\omega$
or $t$). Important universal information is also contained in scaling
functions, which can be used to vary more than one parameter
simultaneously. Here we compute $t^{1-(2-\eta)/z}
\rho(t,0,T_r)$ as a function of time $t$ and reduced temperature $T_r$. According to (\ref{eq:gtTscalingrho}) and (\ref{eq:tcorrelationlength}), this is expected to become only a function of the product $t\, T_r^{\nu z}$ sufficiently close to the critical temperature. In order to use this as well for an independent estimate of the value of $z$, first
we consider results from different temperatures sufficiently close to $T_c$ and
compute the simultaneous best fit taking $z$ as a fit parameter for
known $\eta$ and $\nu$. The least squares fit is based on the integral of 
the local width of the distribution of curves with different temperatures
for a given $ t\, T_r^{\nu z} $. The region of this integration excludes very 
early times showing non-universal behaviour, and only temperatures with less than ten percent deviation from $T_c$ are taken into account. If scaling holds, then a value of $z$ exists
for which all curves at different $T_r$ are on top of each other. This method for extracting $z$ is not very precise, because the quality of the fit turns out not to be very sensitive to changes in $z$. Using this measurement alone, our estimate for the dynamical critical exponent would be $ z= 2.05 \pm 0.15 $, with a somewhat larger error than indicated in (\ref{eq:z}).
\begin{figure}[t]
\begin{center}
\includegraphics[scale=1.,angle=0]{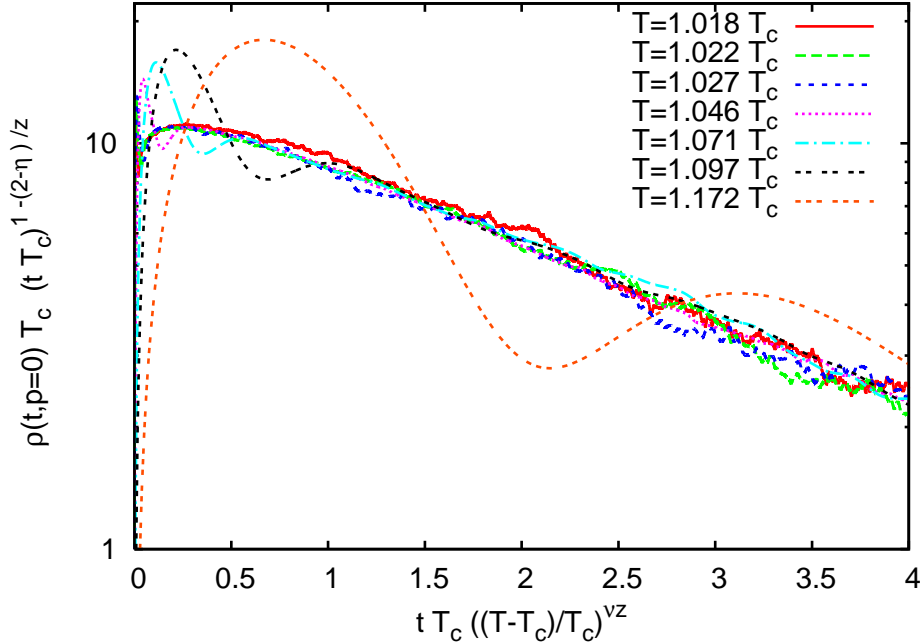}
\caption{Shown is $t^{1-(2-\eta)/z} \rho(t,0,T_r)$ versus $t\, T_r^{\nu z}$ for different temperatures. Sufficiently close to $T_c$ the curves approach a scaling function, which follows an exponential behavior for not too early times in accordance with (\ref{eq:scalingfunction}). 
\label{fig:cuz2.05}}
\end{center}
\end{figure}

\begin{figure}[t]
\begin{center}
\includegraphics[scale=1.0]{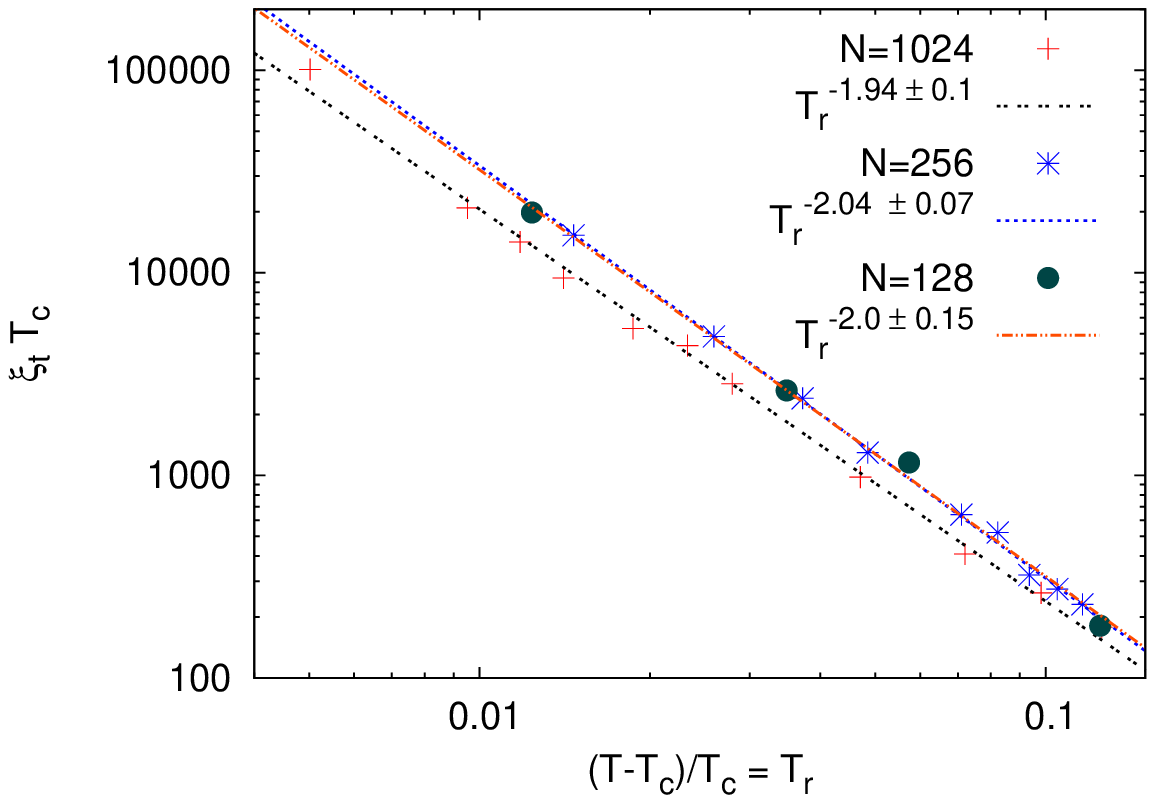}
\caption{The temporal correlation length as a function of the reduced temperature, showing a power-law behavior associated to critical slowing down. \label{fig:ksivst}}
\end{center}
\end{figure}
The result of this procedure is shown in Fig.~\ref{fig:cuz2.05}. 
Though the curves are comparably noisy, which could be reduced 
by increasing statistics at the expense of more computer time, they
nicely exhibit the expected scaling behavior for not too small $t\,
T_r^{\nu z}$ close to $T_c$. After the non-universal short-time physics is damped,
the scaling function exhibits an exponential behavior as a function of
$t\, T_r^{\nu z}$ in accordance with (\ref{eq:scalingfunction}). Indeed, the spectral function becomes well described by $\rho(t,0,T_r) \sim t^{(2-\eta)/z -1}\, \exp (-t/\xi_t) $ in this range. We use this to simultaneously fit the parameters $z$ and $\xi_t$ for each curve corresponding to a given temperature. The result for the temporal correlation length $\xi_t$ as a function of the reduced temperature $T_r$ is shown in Fig.~\ref{fig:ksivst}. It follows a power-law, which is governed by the product of exponents $\nu z$ according to
(\ref{eq:tcorrelationlength}). With $\nu = 1$ we find again a result
agreeing with (\ref{eq:z}) on the few percent accuracy level. While all previous figures show data obtained on a $N=1024$ lattice, Fig.~\ref{fig:ksivst} also includes data measured on smaller lattices. The results indicate that we have reached the infinite volume limit within the given errors.

Going away from the critical domain by sufficiently increasing $T_r$, the physics becomes no longer governed by scaling behavior. To see how the transition to the non-universal regime happens, we have included in Fig.~\ref{fig:cuz2.05} data for the temperature $T = 1.172\, T_c$ that corresponds to the twin-peak spectral function shown in Fig.~\ref{fig:twinpeak} in Fourier space. The non-universal oscillatory behavior visible in Fig.~\ref{fig:cuz2.05} originates from the quasi-particle peak of the spectral function at that temperature, while the oscillation on average still follow the scaling function dependence characteristic for critical phenomena. For even larger deviations from $T_c$, the non-universal dynamics corresponding to a Breit-Wigner like spectral function as discussed above is recovered.    

\begin{figure}[t]
\begin{center}
\includegraphics[scale=1.0]{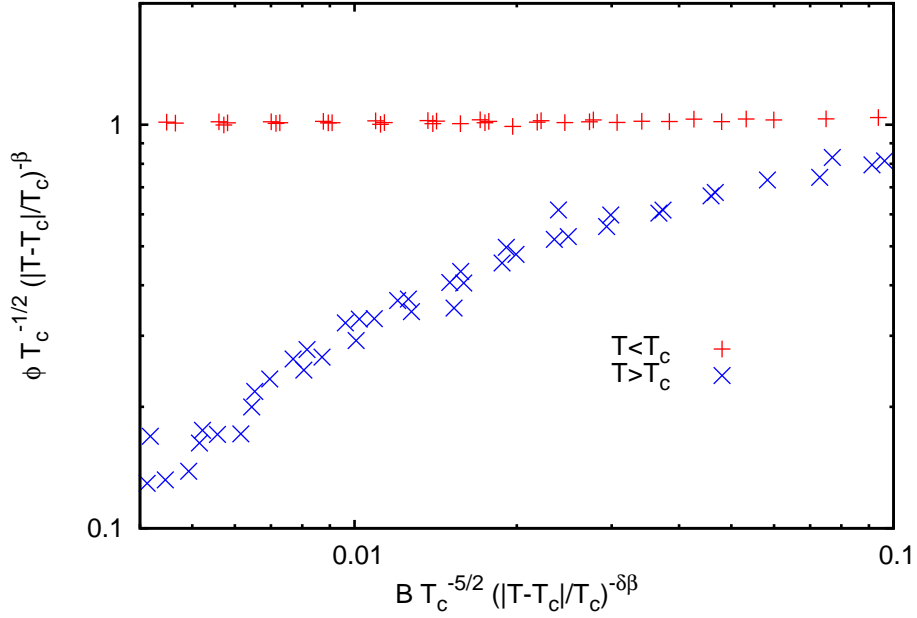}
\caption{Shown is $\phi\, |T_r|^{-\beta}$ versus  $|T_r|^{- \beta \delta }$ for different temperatures and values for the $B$-field using $\beta=1/8$ and $\delta=15$.  
\label{fig:magnetic}}
\end{center}
\end{figure}
So far, we have considered the dynamical information which can be extracted from the time or frequency dependence of two-point functions in thermal equilibrium. A similar analysis can also be performed for higher $n$-point functions. Time translation invariance restricts all $n$-point functions to depend on relative times only. As a consequence, the one-point function is constant, i.e.\ the order parameter does not depend on time in thermal equilibrium. Here we consider the order parameter $\phi(T,B) = \langle \varphi(t,\mathbf{x}) \rangle_{\mathrm{cl}}$ in the presence of a small external field $B$ for temperatures both in the symmetric phase as well as in the phase with spontaneous symmetry breaking. This can be used to verify that our simulations are consistent with the exactly known static exponents for the two-dimensional Ising universality class~\cite{Onsager}. 

The scaling form of the order parameter is given by
\begin{equation}
\phi ( s\, T_r, s^{\beta\delta } B) \,=\, s^{\beta} \phi(T_r,B)  \, .
\end{equation}
Setting $ s\, T_r  = \mathrm{sgn} (T_r) $ one gets
\begin{equation}
\phi(T_r,B) \, = \, |T_r|^{\beta} \phi_\pm (  B |T_r|^{- \beta \delta }  ) \, ,
\label{eq:magnrescale}
\end{equation}
where $\phi_\pm( B |T_r|^{- \beta \delta } ) \equiv \phi (
\mathrm{sgn}(T_r) , B |T_r|^{- \beta \delta } )$ denotes scaling
functions above and below $T_c$, respectively.  Measuring the order
parameter for different temperatures and external field values will
lead to only two curves in the vicinity of $T_c$ if scaling
holds. Using the known values $\beta=1/8$ and $\delta=15$, we indeed
find that all data is well described by two curves with good
accuracy. In turn, one can also use this for known exponents and
unknown value of $T_c$ to obtain an estimate from fitting to the
temperature where the data falls on two curves only. Results on a $N =
128$ lattice are given in Fig.~\ref{fig:magnetic}, which shows
overlapping data from stepping through the temperature range $0.88
\leq T/T_c \leq 1.1$ for a set of ten temperatures with approximately
equal step size and sufficiently small values for the $B$-field. It
nicely demonstrates scaling behavior already on rather small lattices.

\section{Conclusions}

We conclude by putting our results into the context of the standard 
classification scheme of dynamic critical phenomena, which is based on the identification of 
the relevant hydrodynamic degrees of freedom~\cite{Hohenberg:1977ym}.
A low-energy description of the slow dynamics away from $T_c$ requires the inclusion of hydrodynamic modes associated with conservation laws in addition to the order parameter field.
In a single-component real scalar field theory the only locally conserved quantities are energy density and momentum density. In particular, there is no conserved order parameter. The linearized hydrodynamic equations for the relativistic fluid can be reduced to two equations for the transverse part of the momentum density and the local equilibrium pressure~\cite{Jeon:1994if}. The hydrodynamic equation for the transverse part of the momentum density describes diffusive modes with quadratic dispersion relation. In contrast, sound wave modes at sufficiently low momenta and long enough time scales have a linear dispersion relation. As a consequence, they may be effectively integrated out since their frequencies are much higher than the frequencies of the dissipative modes. This means that the variations of pressure caused by sound excitations are fast and average out, such that there are no sound waves on the comparably long diffusive time scales. Therefore, the relevant hydrodynamic mode is the transverse component of the momentum density. According to the classification scheme of Halperin and Hohenberg~\cite{Hohenberg:1977ym}, this low-energy effective theory should be described by the dynamic universality class of relaxational models with conserved density (Model C). For single-component theories in this universality class $z = 2 + \alpha/\nu$, where $\alpha$ denotes the exponent for the singular part of the susceptibility for the conserved density. For $\alpha = 0$ in $d = 2$ up to a logarithmic temperature dependence, one concludes:  
\begin{equation}
z \, = \, 2 \, .
\label{eq:zexpected}
\end{equation} 

Comparison with (\ref{eq:z}) provides a justification from first principles that the relativistic theory for a single-component real scalar field is well described by this dynamic universality class. Some cautionary remarks are in order at this point. For two spatial dimensions, a value of $z$ around two is not very special. For instance, Model A, describing a system with no conservation laws, is known to be characterized by $z \simeq 2.16$ for $d=2$~\cite{daSilva:2008tt}. The present statistics is chosen to be just able to distinguish between the two. For more realistic applications this can be systematically improved, however, at the expense of more computer time. We have mentioned above that a mean field approximation for the critical spectral function would lead to $z=1$ for the relativistic theory. This corresponds to propagating modes with linear dispersion relation. In fact, for temperatures of $T \gtrsim 1.5\, T_c$ the spectral function is found to be rather well approximated by a relativistic Breit-Wigner form with considerable width. We observe the physics of diffusive modes sufficiently close to the phase transition. In the hydrodynamic description this is a consequence of the coupling to a conserved density field.
When the latter is taken properly into account, this consideration would lead to $z=2$ already on a mean field level.  
Our simulations thus demonstrate that the hydrodynamic terms are properly emerging from the underlying microscopic model
in the scaling regime near the phase transition. 
Most strikingly, for a significant intermediate temperature range the spectral function exhibits a twin-peak structure. The strong interplay between propagating relativistic modes and diffusive processes at low energies provides a challenge for effective descriptions in this temperature range. A qualitatively similar behavior would be very interesting in the context of the discussion of the "melting temperature" of states above the QCD transition, if the transition for physical strange quark mass indeed occurs close to a second-order phase transition, or near the QCD critical point for non-zero baryon number density.   

\vspace*{1.0cm}

This work is supported in part by the BMBF grant 06DA267, and by the
DFG under contract SFB634. A large part of the numerical calculations for this project
were done on the bwGRiD (http://www.bw-grid.de), member of the German D-Grid
initiative, funded by the 
{\em Bundesministerium f{\"u}r Bildung und Forschung} and the {\em Ministerium f{\"u}r
Wissenschaft, Forschung und Kunst Baden-W{\"u}rttemberg}.

\end{document}